\documentclass[runningheads]{llncs}
\usepackage[T1]{fontenc}

\usepackage{graphicx}
\usepackage{makeidx}  
\usepackage{graphicx}
\usepackage{subfigure}
\usepackage{multirow}
\usepackage{color,soul} 
\usepackage{ dsfont }
\usepackage{amssymb,amsfonts}
\usepackage{physics} 
\usepackage{cases}
\usepackage{hyperref}
\usepackage{multirow}
\usepackage{amsmath}

\begin{document}
\title{An AI-Ready Multiplex Staining Dataset for Reproducible and Accurate Characterization of Tumor Immune Microenvironment}
\titlerunning{AI-Ready Multiplex Staining Dataset}
%
\author{Parmida Ghahremani\inst{1}* \and Joseph Marino\inst{1}* \and Juan Hernandez-Prera\inst{2} \and Janis V. de la Iglesia\inst{2} \and Robbert JC Slebos\inst{2} \and Christine H. Chung\inst{2}* \and Saad Nadeem\inst{1}*}

\authorrunning{Ghahremani et al.}

\institute{Memorial Sloan Kettering Cancer Center, New York NY 10065, USA \and
Moffitt Cancer Center, Tampa FL 33612, USA\\
*Equal contribution. Email: \email{christine.chung@moffitt.org} and \email{nadeems@mskcc.org}
}

\maketitle              
\begin{abstract}
We introduce a new AI-ready computational pathology dataset containing restained and co-registered digitized images from eight head-and-neck squamous cell carcinoma patients. Specifically, the same tumor sections were stained with the expensive multiplex immunofluorescence (mIF) assay first and then restained with cheaper multiplex immunohistochemistry (mIHC). This is a first public dataset that demonstrates the equivalence of these two staining methods which in turn allows several use cases; due to the equivalence, our cheaper mIHC staining protocol can offset the need for expensive mIF staining/scanning which requires highly-skilled lab technicians. As opposed to subjective and error-prone immune cell annotations from individual pathologists (disagreement $>$ 50\%) to drive SOTA deep learning approaches, this dataset provides objective immune and tumor cell annotations via mIF/mIHC restaining for more reproducible and accurate characterization of tumor immune microenvironment (e.g. for immunotherapy). We demonstrate the effectiveness of this dataset in three use cases: (1) IHC quantification of CD3/CD8 tumor-infiltrating lymphocytes via style transfer, (2) virtual translation of cheap mIHC stains to more expensive mIF stains, and (3) virtual tumor/immune cellular phenotyping on standard hematoxylin images. The dataset is available at \url{https://github.com/nadeemlab/DeepLIIF}.

\end{abstract}

\section{Introduction}
Accurate spatial characterization of tumor immune microenvironment is critical for precise therapeutic stratification of cancer patients (e.g. via immunotherapy). Currently, this characterization is done manually by individual pathologists on standard hematoxylin-and-eosin (H\&E) or singleplex immunohistochemistry (IHC) stained images. However, this results in high interobserver variability among pathologists, primarily due to the large ($>$ 50\%) disagreement among pathologists for immune cell phenotyping \cite{reisenbichler2020prospective}. This is also a big cause of concern for publicly available H\&E/IHC cell segmentation datasets with immune cell annotations from single pathologists. Multiplex  staining resolves this issue by allowing different tumor and immune cell markers to be stained on the same tissue section, avoiding any phenotyping guesswork from pathologists.

\begin{figure*}[t!]
    \centering
    \includegraphics[width=\textwidth]{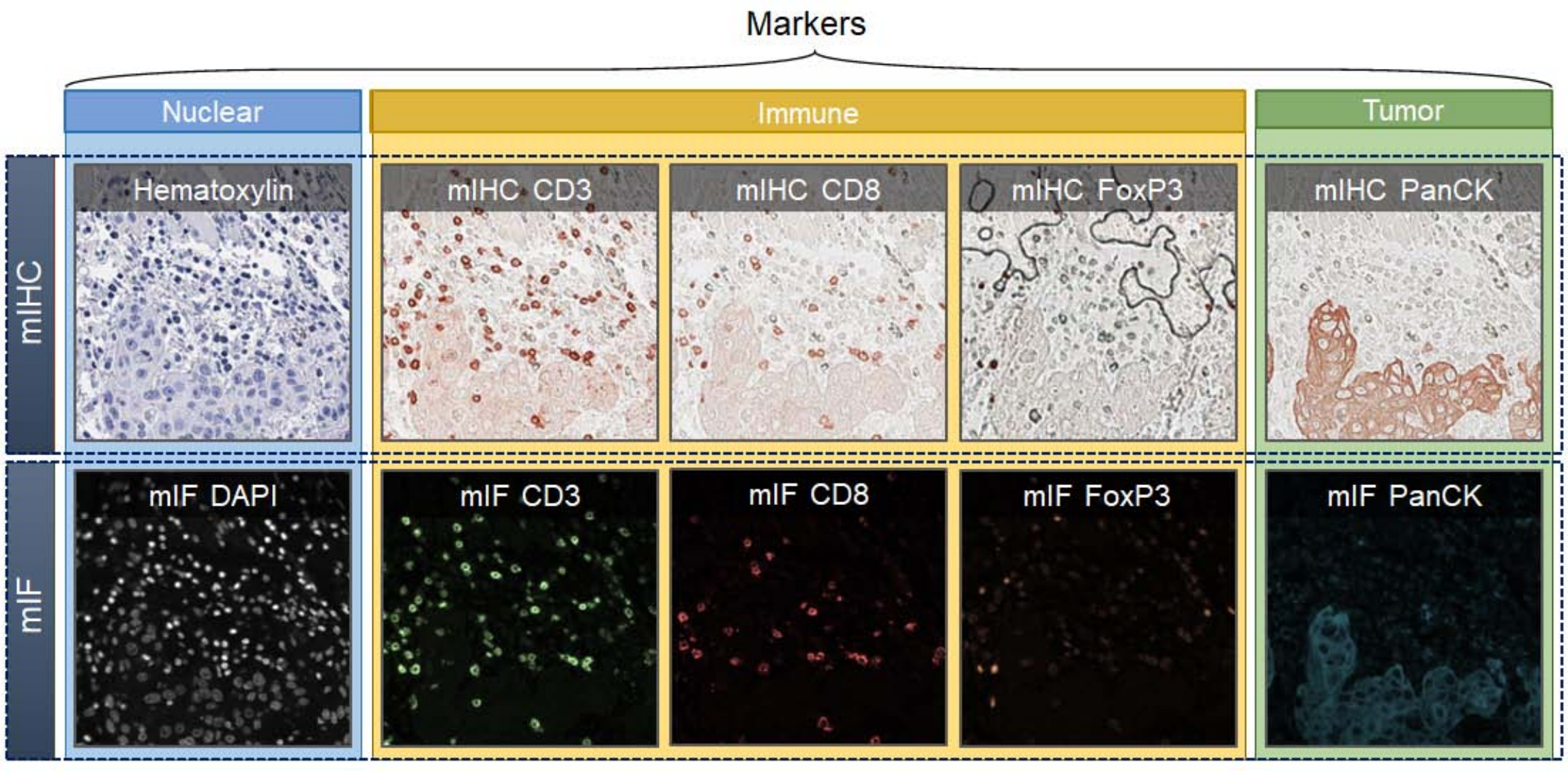}
    \caption{Dataset overview. Restained and co-registered mIHC and mIF sample with nuclear (hematoxylin/DAPI), immune (CD3 - T-cell marker, CD8 - Cytotoxic T-cell, FoxP3 - regulatory T-cell), and tumor (PanCK) markers. CD3 = CD8 + FoxP3.}
    \label{fig:dataset_sample}
\end{figure*}

Multiplex staining can be performed using expensive multiplex immunofluorescence (mIF) or via cheaper multiplex immunohistochemistry (mIHC) assays. MIF staining (requiring expensive scanners and highly skilled lab technicians) allows multiple markers to be stained/expressed on the same tissue section (no co-registration needed) while also providing the utility to turn ON/OFF individual markers as needed. In contrast, current brightfield mIHC staining protocols relying on DAB (3,3'-Diaminobenzidine) alcohol-insoluble chromogen, even though easily implementable with current clinical staining protocols, suffer from occlusion of signal from sequential staining of additional markers. To this effect, we introduce a new brightfield mIHC staining protocol using alcohol-soluble aminoethyl carbazole (AEC) chromogen which allows repeated stripping, restaining, and scanning of the same tissue section with multiple markers. This requires only affine registration to align the digitized restained images to obtain non-occluded signal intensity profiles for all the markers, similar to mIF staining/scanning.

In this paper, we introduce a new dataset that can be readily used out-of-the-box with any artificial intelligence (AI)/deep learning algorithms for spatial characterization of tumor immune microenvironment and several other use cases. To date, only two denovo stained datasets have been released publicly: BCI H\&E and singleplex IHC HER2 dataset \cite{liu2022bci} and DeepLIIF singleplex IHC Ki67 and mIF dataset \cite{ghahremani2022deep}, both without any immune or tumor markers. In contrast, we release the first denovo mIF/mIHC stained dataset with tumor and immune markers for more accurate characterization of tumor immune microenvironment. We also demonstrate several interesting use cases: (1) IHC quantification of CD3/CD8 tumor-infiltrating lymphocytes (TILs) via style transfer, (2) virtual translation of cheap mIHC stains to more expensive mIF stains, and (3) virtual tumor/immune cellular phenotyping on standard hematoxylin images.

\section{Dataset}
The complete staining protocols for this dataset are given in the accompanying \textbf{supplementary material}. Images were acquired at 20$\times$ magnification at Moffitt Cancer Center. The demographics and other relevant information for all eight head-and-neck squamous cell carcinoma patients is given in Table \ref{tab:demographics}.

\begin{table*}[t!]
    \caption{Demographics and other relevant details of the eight anonymized head-and-neck squamous cell carcinoma patients, including ECOG performance score, Pack-Year, and surgical pathology stage (AJCC8).}
    \centering
    \begin{tabular}{|c|c|c|c|c|c|c|c|c|c|}
        \hline
        ID & Age & Gender & Race & ECOG & Smoking & PY & pStage & Cancer Site & Cancer Subsite \\
        \hline
        Case1 & 49 & Male & White & 3 & Current & 21 & 1 & Oral Cavity	& Ventral Tongue \\
        Case2 & 64 & Male & White & 3 & Former & 20 & 4 & Larynx	& Vocal Cord \\
        Case3 & 60 & Male & Black & 2 & Current & 45 & 4 & Larynx	& False Vocal Cord \\
        Case4 & 53 & Male & White & 1 & Current & 68 & 4 & Larynx	& Supraglottic \\
        Case5 & 38 & Male & White & 0 & Never & 0 & 4 & Oral Cavity	& Lateral Tongue \\
        Case6 & 76 & Female & White & 1 & Former & 30 & 2 & Oral Cavity & Lateral Tongue \\
        Case7 & 73 & Male & White & 1 & Former & 100 & 3 & Larynx & Glottis \\
        Case8 & 56 & Male & White & 0 & Never & 0 & 2 & Oral Cavity & Tongue\\
        \hline
    \end{tabular}
    \label{tab:demographics}
\end{table*}

\subsection{Region-of-interest selection and image registration}
After scanning the full images at low resolution, nine regions of interest (ROIs) from each slide were chosen by an experienced pathologist on both mIF and mIHC images: three in the tumor core (TC), three at the tumor margin (TM), and three outside in the adjacent stroma (S) area. The size of the ROIs was standardized at 1356$\times$1012 pixels with a resolution of 0.5 $\mu$m/pixel for a total surface area of 0.343 mm$^2$. Hematoxylin-stained ROIs were first used to align all the mIHC marker images in the open source Fiji software using affine registration. After that, hematoxylin- and DAPI-stained ROIs were used as references to align mIHC and mIF ROIs again using Fiji and subdivided into 512$\times$512 patches, resulting in total of 268 co-registered mIHC and mIF patches ($\sim$33 co-registered mIF/mIHC images per patient).

\subsection{Concordance study}
\begin{figure*}[t!]
    \centering
    \includegraphics[width=\textwidth]{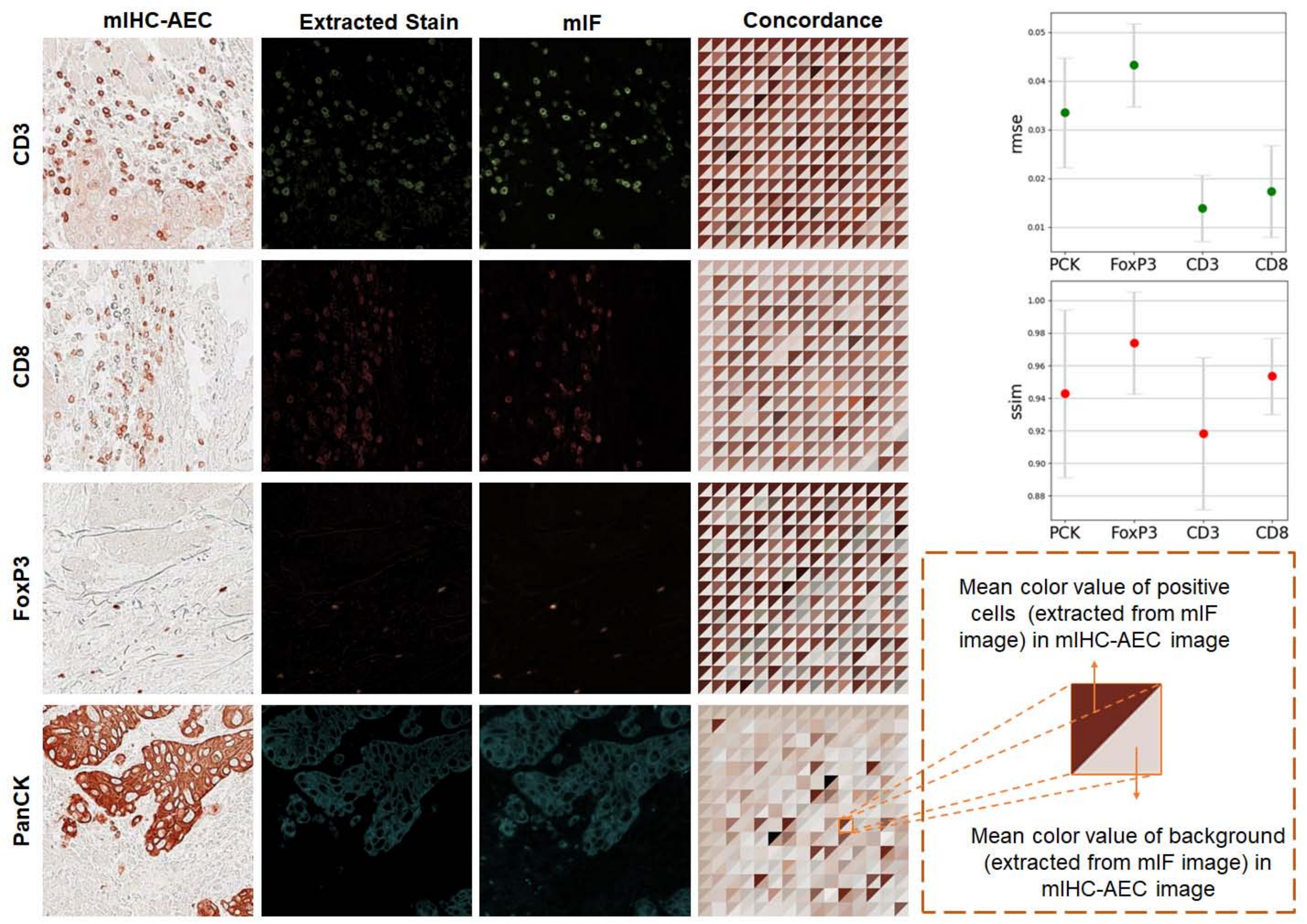}
    \caption{Concordance Study. Second column shows stains extracted from first column mIHC-AEC images using Otsu thresholding. Third column shows the corresponding perfectly co-registered original mIF images. Using the mIF image, we separated foreground of the mIHC-AEC image from its background and calculated the mean value of the foreground pixels as well as the background pixels. The fourth column shows the results of the concordance study. Each square represents an image in the dataset and the top half of each square shows the mean color value of the positive cells, extracted from mIHC-AEC using its corresponding mIF image and the bottom half of it shows the mean color value of its background. \textit{The high intensity of the top half of the squares represents positive cells and the low intensity of the bottom half represents non-positive cells (background), which is seen in almost all squares, demonstrating \textbf{high concordance among mIHC-AEC and mIF data}.} The last column shows the RMSE and SSIM diagrams of all four stains calculated using the extracted stain from IHC-AEC images (second column) and the mIF images (third column). The low error rate of RMSE and high structural similarity seen in these diagrams show high concordance among mIHC-AEC and mIF images.}
    \label{fig:concordance_study}
\end{figure*}

We compared mIF and mIHC assays for concordance in marker intensities. The results are shown in Figure \ref{fig:concordance_study}. This is the first direct comparison of mIF and mIHC using identical slides. It provides a standardized dataset to demonstrate the equivalence of the two methods and a source that can be used to calibrate other methods. 

\begin{figure*}[t]
    \centering
    \includegraphics[width=0.8\textwidth]{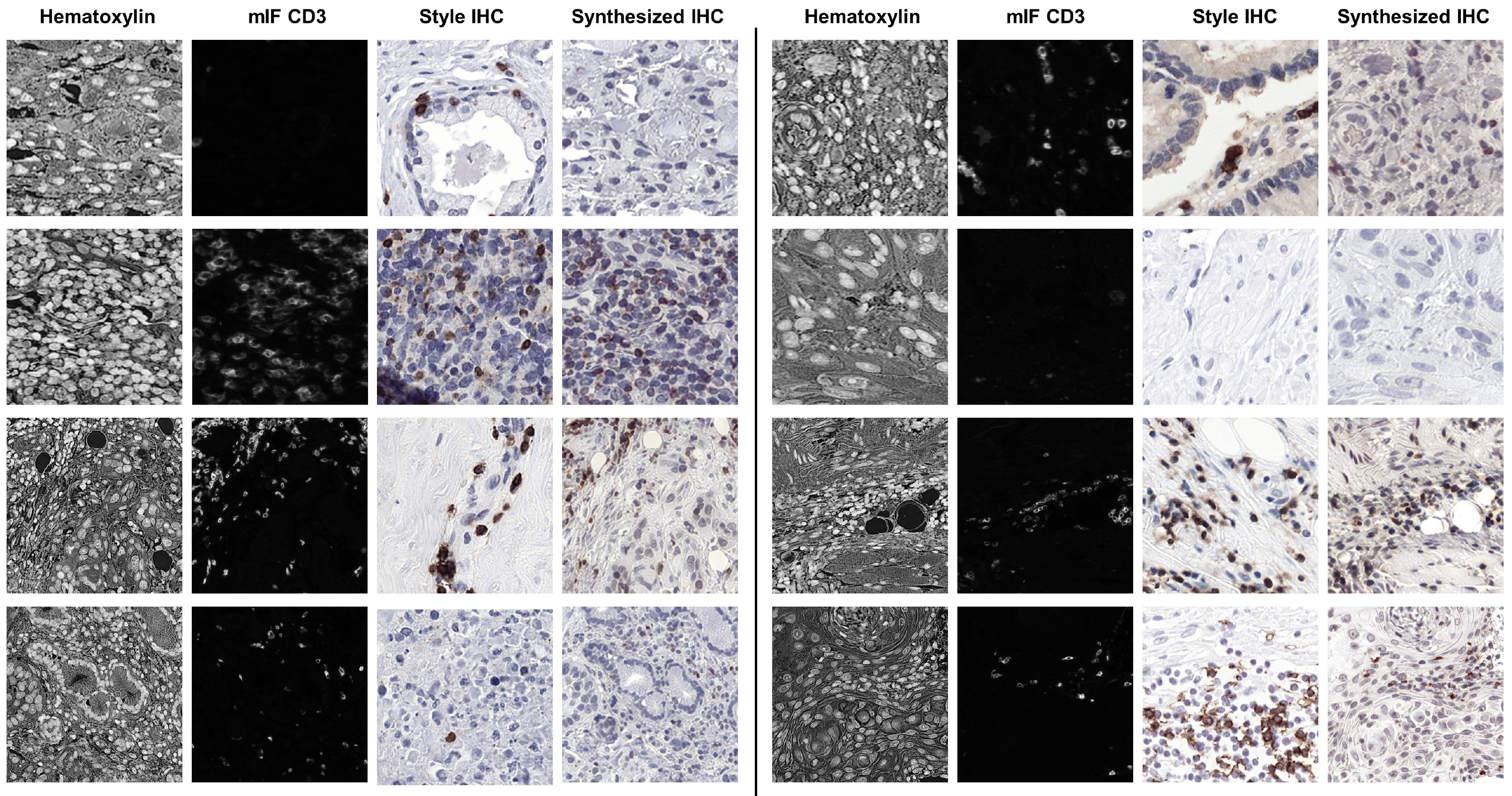}
    \caption{Examples of synthesized IHC images and corresponding input images. Style IHC images were taken from the public LYON19 challenge dataset~\cite{swiderska2019learning}. We used grayscale Hematoxylin images because they performed better with style transfer.}
    \label{fig:style_results}
\end{figure*}

\begin{figure*}[h!]
    \centering
    \includegraphics[width=0.8\textwidth]{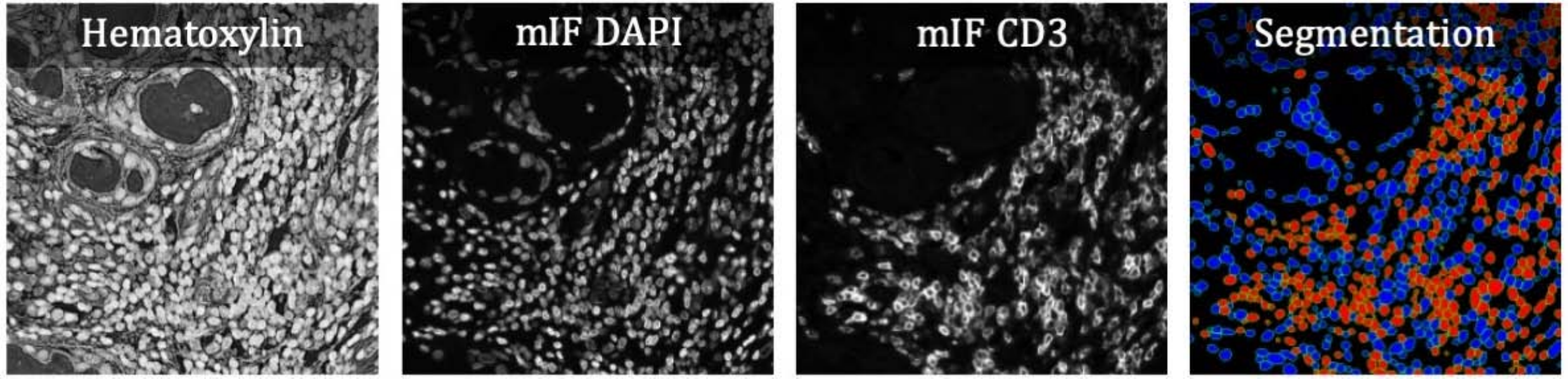}
    \caption{Examples of Hematoxylin, mIF DAPI, mIF CD3 and classified segmentation mask for this marker. The DAPI images were segmented using Cellpose~\cite{stringer2021cellpose} and manually corrected by a trained technician and approved by a pathologist. The segmented masks were classified using the CD3 channel intensities.}
    \label{fig:seg_dataset_figure}
\end{figure*}

\section{Use cases}
In this section, we demonstrate some of the use cases enabled by this high-quality AI-ready dataset. We have used publicly available state-of-the-art tools such as Adaptive Attention Normalization (AdaAttN)~\cite{liu2021adaattn} for style transfer in the IHC CD3/CD8 quantification use case and DeepLIIF virtual stain translation \cite{ghahremani2022deep,ghahremani2022deepliif} in the remaining two use cases.

\begin{table*}[t!]
\centering
\caption{Quantitative metrics for NuClick and LYSTO testing sets. \textbf{F1} is the harmonic mean of recall and precision, \textbf{IOU} is intersection over union, and pixel accuracy (PixAcc) is $\frac{TP}{TP+FP+FN}$, where TP, FP, and FN represent the number of true positive, false positive, and false negative pixels, respectively.}
\label{tab:results}
\setlength{\tabcolsep}{4pt}
\centering
\begin{tabular}{l|c|c|c|c|c}
\hline
& & \multicolumn{3}{|c|}{\textbf{NuClick}} & \textbf{LYSTO}\\
\hline
\textbf{Model} & \textbf{Dataset} & \textbf{F1}$\uparrow$ & \textbf{IOU}$\uparrow$ & \textbf{PixAcc}$\uparrow$ & \textbf{DiffCount}$\downarrow$\\
\hline \hline
\multirow{2}{*}{UNet~\cite{ronneberger2015u}} & NuClick & 0.47 $\pm$ 0.30 & 0.36 $\pm$ 0.24 & 0.62 $\pm$ 0.37 & 10.06 $\pm$ 15.69\\
\cline{2-6} 
& Our Dataset & 0.48 $\pm$ 0.29 & 0.36 $\pm$ 0.25 & 0.69 $\pm$ 0.37 & 2.91 $\pm$ 5.47\\
\hline
\hline
\multirow{2}{*}{FPN~\cite{kirillov2017unified}} & NuClick & 0.50 $\pm$ 0.31 & 0.39 $\pm$ 0.26 & 0.64 $\pm$ 0.38 & 2.82 $\pm$ 3.49\\
\cline{2-6} 
& Our Dataset & 0.52 $\pm$ 0.31 & 0.40 $\pm$ 0.26 & 0.67 $\pm$ 0.36 & 1.90 $\pm$ 2.90\\
\hline
\hline
\multirow{2}{*}{UNet++~\cite{zhou2018unet++}} & NuClick & 0.49 $\pm$ 0.30 & 0.37 $\pm$ 0.25 & 0.63 $\pm$ 0.37 & 2.75 $\pm$ 5.29\\
\cline{2-6} 
& Our Dataset & 0.53 $\pm$ 0.30 & 0.41 $\pm$ 0.26 & 0.70 $\pm$ 0.36 & 2.19 $\pm$ 2.89\\
\hline

\hline
\end{tabular}
\end{table*}

\subsection{IHC CD3/CD8 scoring using mIF style transfer}
We generate a stylized IHC image (Figure \ref{fig:style_results}) using three input images: (1) hematoxylin image (used for generating the underlying structure of cells in the stylized image), (2) its corresponding mIF CD3/CD8 marker image (used for staining positive cells as brown), and (3) sample IHC style image (used for transferring its style to the final image). The complete architecture diagram is given in the \textbf{supplementary material}. Specifically, the model consists of two sub-networks: \\
\textbf{(a) Marker Generation:} This sub-network is used for generating mIF marker data from the generated stylized image. We use a conditional generative adversarial network (cGAN) \cite{isola2017image} for generating the marker images. The cGAN network consists of a generator, responsible for generating mIF marker images given an IHC image, and a discriminator, responsible for distinguishing the output of the generator from ground truth data. We first extract the brown (DAB channel) from the given style IHC image, using stain deconvolution. Then, we use pairs of the style images and their extracted brown DAB marker images to train this sub-network. This sub-network improves staining of the positive cells in the final stylized image by comparing the extracted DAB marker image from the stylized image and the input mIF marker image at each iteration.\\
\textbf{(b) Style Transfer:} This sub-network creates the stylized IHC image using an attention module, given (1) the input hematoxylin and the mIF marker images and (2) the style and its corresponding marker images. For synthetically generating stylized IHC images, we follow the approach outlined in AdaAttN~\cite{liu2021adaattn}. We use a pre-trained VGG-19 network~\cite{simonyan2014very} as an encoder to extract multi-level feature maps and a decoder with a symmetric structure of VGG-19. We then use both shallow and deep level features by using AdaAttN modules on multiple layers of VGG. This sub-network is used to create a stylized image using the structure of the given hematoxylin image while transferring the overall color distribution of the style image to the final stylized image. The generated marker image from the first sub-network is used for a more accurate colorization of the positive cells against the blue hematoxylin counterstain/background; not defining loss functions based on the markers generated by the first sub-network leads to discrepancy in the final brown DAB channel synthesis.

For the stylized IHC images with ground truth CD3/CD8 marker images, we also segmented corresponding DAPI images using our interactive deep learning ImPartial~\cite{martinez2021impartial} tool \url{https://github.com/nadeemlab/ImPartial} and then classified the segmented masks using the corresponding CD3/CD8 channel intensities, as shown in Figure \ref{fig:seg_dataset_figure}. We extracted 268 tiles of size $512 \times 512$ from this final segmented and co-registered dataset. For the purpose of training and testing all the models, we extract four images of size $256\times 256$ from each tile due to the size of the external IHC images, resulting in a total of 1072 images. We randomly extracted tiles from the LYON19 challenge dataset~\cite{swiderska2019learning} to use as style IHC images. Using these images, we created a dataset of synthetically generated IHC images from the hematoxylin and its marker image as shown in Figure~\ref{fig:style_results}. 


\begin{figure*}[t!]
    \centering
    \includegraphics[width=0.8\textwidth]{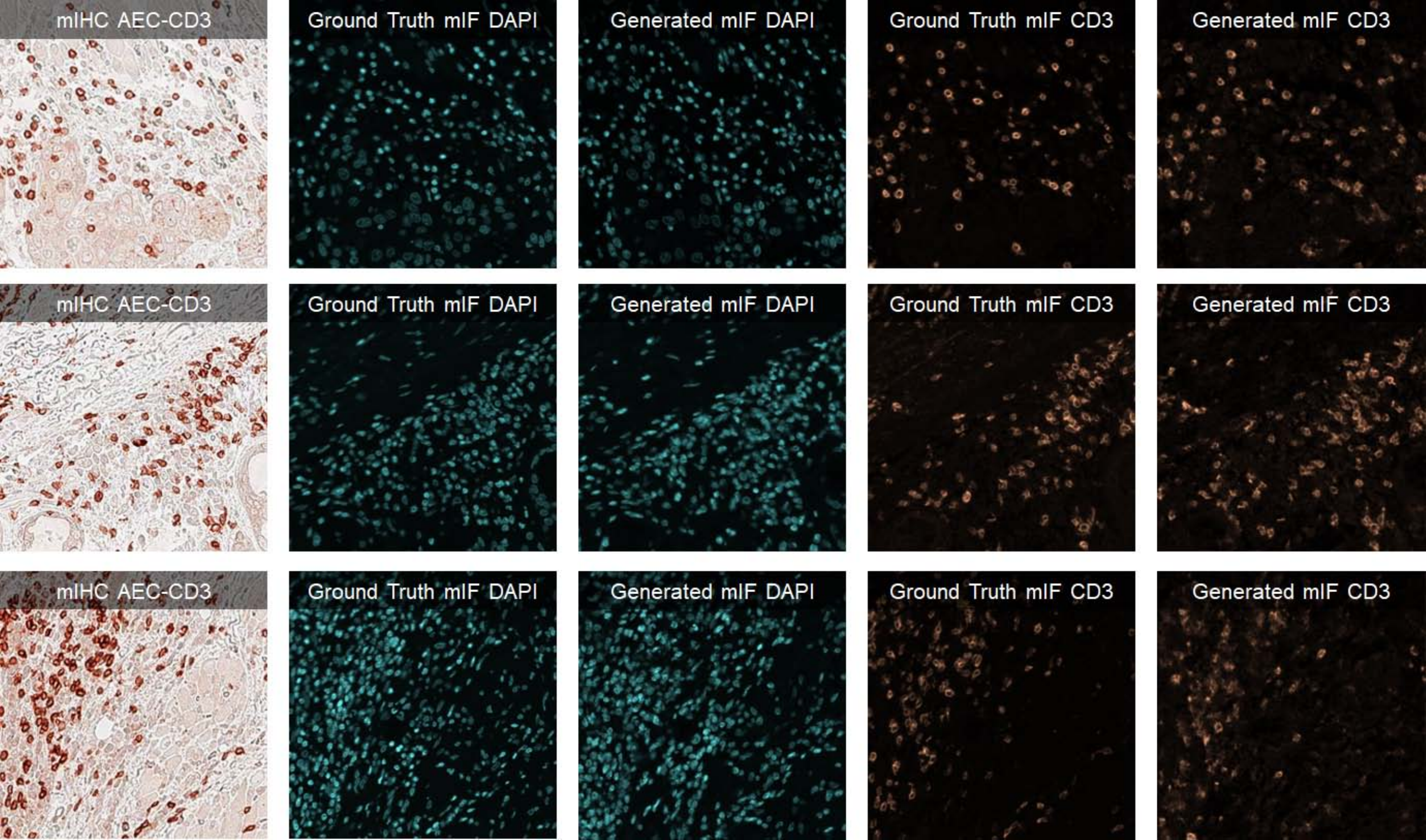}
    \caption{Examples of ground-truth and generated mIF data from mIHC-AEC images.}
    \label{fig:results_aec_mpif}
\end{figure*}

We evaluated the effectiveness of our synthetically generated dataset (stylized IHC images and corresponding segmented/classified masks) using our generated dataset with the NuClick training dataset (containing manually segmented CD3/CD8 cells)~\cite{koohbanani2020nuclick}. We randomly selected 840 and 230 patches of size $256 \times 256$ from the created dataset for training and validation, respectively. NuClick training and validation sets \cite{koohbanani2020nuclick} comprise 671 and 200 patches, respectively, of size $256 \times 256$ extracted from LYON19 dataset~\cite{swiderska2019learning}. LYON19 IHC CD3/CD8 images are taken from breast, colon, and prostate cancer patients. We split their training set into training and validation sets, containing 553 and 118 images, respectively, and use their validation set for testing our trained models. We trained three models including UNet~\cite{ronneberger2015u}, FPN~\cite{kirillov2017unified}, UNet++~\cite{zhou2018unet++} with the backbone of resnet50 for 200 epochs and early stopping on validation score with patience of 30 epochs, using binary cross entropy loss and Adam optimizer with learning rate of 0.0001. As shown in Table~\ref{tab:results}, models trained with our synthetic training set outperform those trained solely with NuClick data in all metrics. 

We also tested the trained models on 1,500 randomly selected images from the training set of the Lymphocyte Assessment Hackathon (LYSTO)~\cite{ciompi2019lymphocyte}, containing image patches of size $299\times 299$ obtained at a magnification of $40\times$ from breast, prostate, and colon cancer whole slide images stained with CD3 and CD8 markers. Only the total number of lymphocytes in each image patch are reported in this dataset. To evaluate the performance of trained models on this dataset, we counted the total number of marked lymphocytes in a predicted mask and calculated the difference between the reported number of lymphocytes in each image with the total number of lymphocytes in the predicted mask by the model. In Table~\ref{tab:results}, the average difference value (\textbf{DiffCount}) of lymphocyte number for the whole dataset is reported for each model. As seen, the trained models on our dataset outperform the models trained solely on NuClick data.

\begin{figure*}[t]
    \centering
    \includegraphics[width=0.8\textwidth]{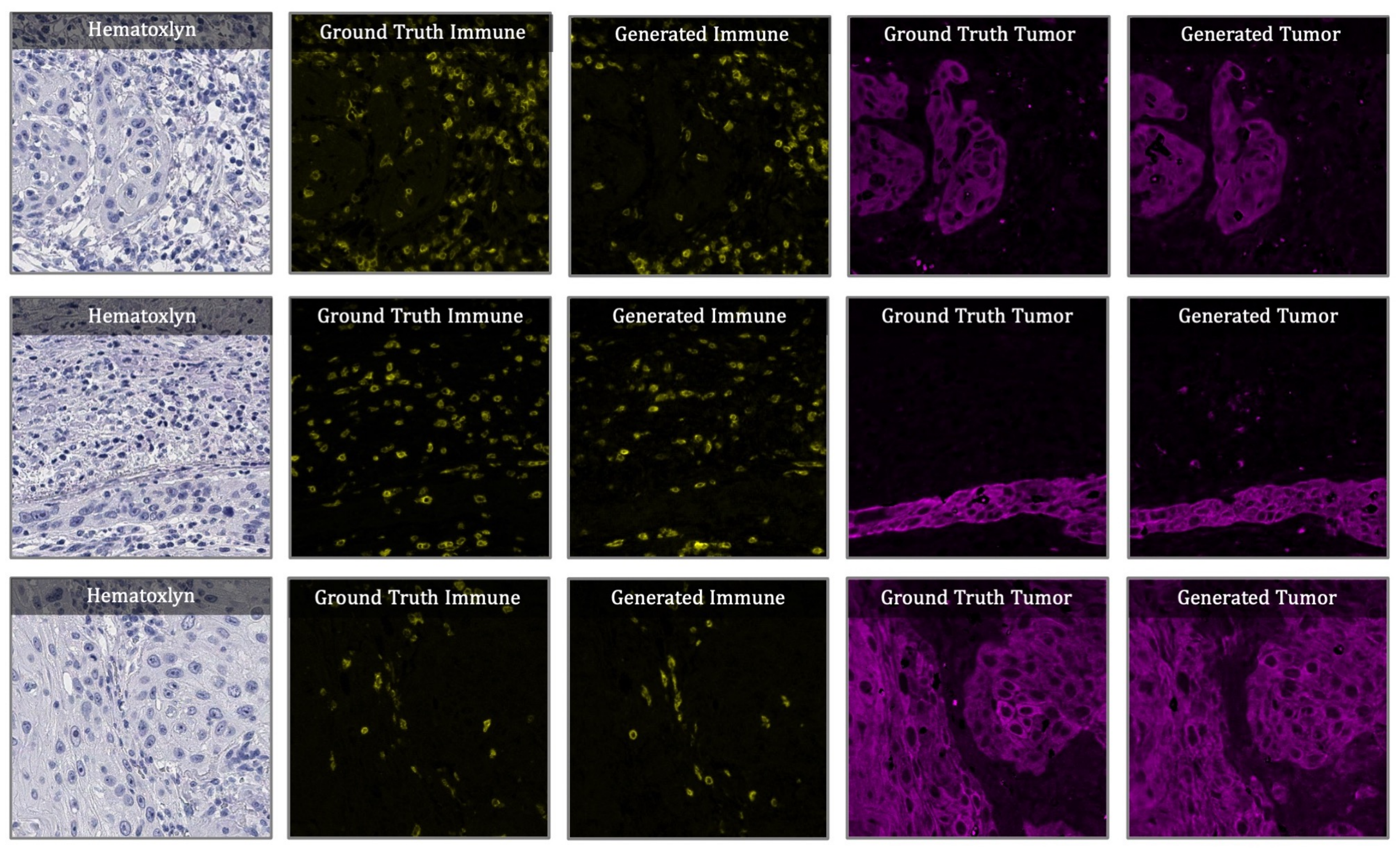}
    \caption{Examples of ground-truth and generated mIF immune (CD3) and tumor (PanCK) markers from standard hematoxylin images.}
    \label{fig:results_hema_immune_tumor}
\end{figure*}

\subsection{Virtual translation of cheap mIHC to expensive mIF stains}
Unlike clinical DAB staining, as shown in style IHC images in Figure \ref{fig:style_results}, where brown marker channel has a blue hematoxylin nuclear counterstain to stain for all the cells, our mIHC AEC-stained marker images (Figure \ref{fig:results_aec_mpif}) do not stain for all the cells including nuclei. In this use case, we show that mIHC marker images can be translated to higher quality mIF DAPI and marker images which stain effectively for all the cells. We used the publicly available DeepLIIF virtual translation module \cite{ghahremani2022deep,ghahremani2022deepliif} for this task. We trained DeepLIIF on mIHC CD3 AEC-stained images to infer mIF DAPI and CD3 marker. Some examples of testing the trained model on CD3 images are shown in Figure \ref{fig:results_aec_mpif}. We calculated the Mean Squared Error (MSE) and Structural Similarity Index (SSIM) to evaluate the quality of the inferred modalities by the trained model. The MSE and SSIM for mIF DAPI was 0.0070 and 0.9991 and for mIF CD3 was 0.0021 and 0.9997, indicating high accuracy of mIF inference.

\subsection{Virtual cellular phenotyping on standard hematoxylin images}
There are several public H\&E/IHC cell segmentation datasets with manual immune cell annotations from single pathologists. These are highly problematic given the large ($>$ 50\%) disagreement among pathologists on immune cell phenotyping \cite{reisenbichler2020prospective}. In this last use case, we infer immune and tumor markers from the standard hematoxylin images using again the public DeepLIIF virtual translation module \cite{ghahremani2022deep,ghahremani2022deepliif}. We train the translation task of DeepLIIF model using the hematoxylin, immune (CD3) and tumor (PanCK) markers. Sample images/results taken from the testing dataset are shown in Figure \ref{fig:results_hema_immune_tumor}.

\section{Conclusions and Future Work} 
We have released the first AI-ready restained and co-registered mIF and mIHC dataset for head-and-neck squamous cell carcinoma patients. This dataset can be used for virtual phenotyping given standard clinical hematoxylin images, virtual clinical IHC DAB generation with ground truth segmentations (to train high-quality segmentation models across multiple cancer types) created from cleaner mIF images, as well as for generating standardized clean mIF images from neighboring H\&E and IHC sections for registration and 3D reconstruction of tissue specimens. In the future, we will release similar datasets for additional cancer types as well as release for this dataset corresponding whole-cell segmentations via ImPartial \url{https://github.com/nadeemlab/ImPartial}.

\noindent \textbf{Data use declaration and acknowledgment:} This study is not Human Subjects Research because it was a secondary analysis of results from biological specimens that were not collected for the purpose of the current study and for which the samples were fully anonymized. This work was supported by MSK Cancer Center Support Grant/Core Grant (P30 CA008748) and by James and Esther King Biomedical Research Grant (7JK02) and Moffitt Merit Society Award to C. H. Chung. It is also supported in part by the Moffitt’s Total Cancer Care Initiative, Collaborative Data Services, Biostatistics and Bioinformatics, and Tissue Core Facilities at the H. Lee Moffitt Cancer Center and Research Institute, an NCI-designated Comprehensive Cancer Center (P30-CA076292). 

%
%
%




%
\title{An AI-Ready Multiplex Staining Dataset for Reproducible and Accurate Characterization of Tumor Immune Microenvironment (Supplement)}
\titlerunning{AI-Ready Multiplex Staining Dataset}
%
\author{Parmida Ghahremani\inst{1}* \and Joseph Marino\inst{1}* \and Juan Hernandez-Prera\inst{2} \and Janis V. de la Iglesia\inst{2} \and Robbert JC Slebos\inst{2} \and Christine H. Chung\inst{2}* \and Saad Nadeem\inst{1}*}

\authorrunning{Ghahremani et al.}

\institute{Memorial Sloan Kettering Cancer Center, New York NY 10065, USA \and
Moffitt Cancer Center, Tampa FL 33612, USA\\
*Equal Contribution. Email: \email{Christine.Chung@moffitt.org} and \email{nadeems@mskcc.org}
}
\maketitle              

\section{Staining Protocols}\label{sec:dataset}

\subsection{Multiplex immunofluorescence assay using multispectral microscopy}
Multiplex immunofluorescence assay using multispectral microscopy
Antibodies against the following were used for staining: CD3 (clone F7.2.38, dilution 1:100, Dako, Carpinteria, CA), CD8 (clone C8/144B, dilution 1:100, Dako, Carpinteria, CA), FOXP3 (clone 236A/E7, dilution 1:200, Abcam, Cambridge, MA) and pancytokeratin (clone AE1/AE3, dilution 1:600, Dako, Carpinteria, CA) and DAPI for nuclei. Following the manufacturer’s instructions, Opal 7-colors kit (same order than antibodies) Opal520 (green), Opal 540 (pink), Opal 570 (yellow), Opal 620 (Orange), Opal 650 (red), and Opal 690 (Aqua blue) was used the dilution 1:150. The Opal 7-colors kit (Opal 520, Opal 540, Opal 570, Opal 620, Opal 650, and Opal 690) was used according to manufacturer’s instructions (PerkinElmer, Waltham, MA). A positive control was used for each protein as follows: human tonsil for CD3 and CD8, FOXP3, and pancytokeratin AE1/AE3. A negative control slide for autofluorescence was included and stained with primary and secondary antibody but omitting fluorochrome-tyramide. The multiplex staining was imaged by using the fluorescence protocol at 10 nm from 420 nm to 720 nm, to extract fluorescence intensity information from the images. For the multiplex immunofluorescence slides, each set of slides was scanned with the Vectra imaging system.

\subsection{Multiplex immunohistochemistry assay using high resolution imaging}
The identical slides used for mIF were processed for mIHC by removing the coverslips and rehydration in gradient ethanol series three times in 100\%, 90\%, 70\% and water at 5 min each. Nuclei were stained using Hematoxylin (\#SH26, Fisher Scientific, Pittsburgh, PA) and stained slides were covered with micro cover glass (\#48393081, VWR, Radnor, PA) using distilled water to prevent tissue drying and Whole Slide Images (WSI) were created by scanning the slides using a digital pathology scanner Aperio CS2 (Leica Biosystems, Buffalo Grove, IL). Slides were destained from hematoxylin by washing with 1\% HCL for 1 min, followed by antigen retrieval with citrate buffer (pH=6.0) or EDTA (pH=8.0) (\#AB93678 and \#AB93678, respectively, Abcam, Cambridge, MA), ), using a high-pressure cooker (95-100◦C) for 30 min. The slides were incubated in 3\% hydrogen peroxide (\#H325-500, Fisher Scientific, Pittsburgh, PA) twice for 15 min each to block endogenous peroxidase activity followed by serum (\#55984, MP Biomedicals, Santa Ana, CA) incubation for 15 min at room temperature (RT). The first primary antibody was applied for 60 min at RT or 4oC overnight followed by three washes using wash buffer (\#K8007, Dako, Troy, MI). The slides were incubated with appropriate secondary universal immuno-peroxidase polymer, anti-rabbit  or anti-mouse antibody (\#414141F or \#414131F, respectively, Nichirei Bioscience, Tokyo, Japan) for 30 min, followed by three washes and the chromogenic signal was detected using AEC (\#415184F, Nichirei Bioscience, Tokyo, Japan). WSI scans were created using the Aperio CS2 scanner as described above. For subsequent staining with the next antibody, slides were washed in gradient ethanol (70\%, 90\%, 100\%; 2 min each and 100\%, 90\%, 70\%, water; 5 min each], followed by antigen retrieval for 30 min and serum blocking 15 min and the above staining procedure was repeated for each antibody incubation. The order of staining was: Hematoxylin, FoxP3, CD8, CD3, and PanCK; weak stains were performed first since these are easily washed out and are more sensitive to multiple rounds of stripping.

\begin{figure*}[t]
    \centering
    \includegraphics[width=0.8\textwidth]{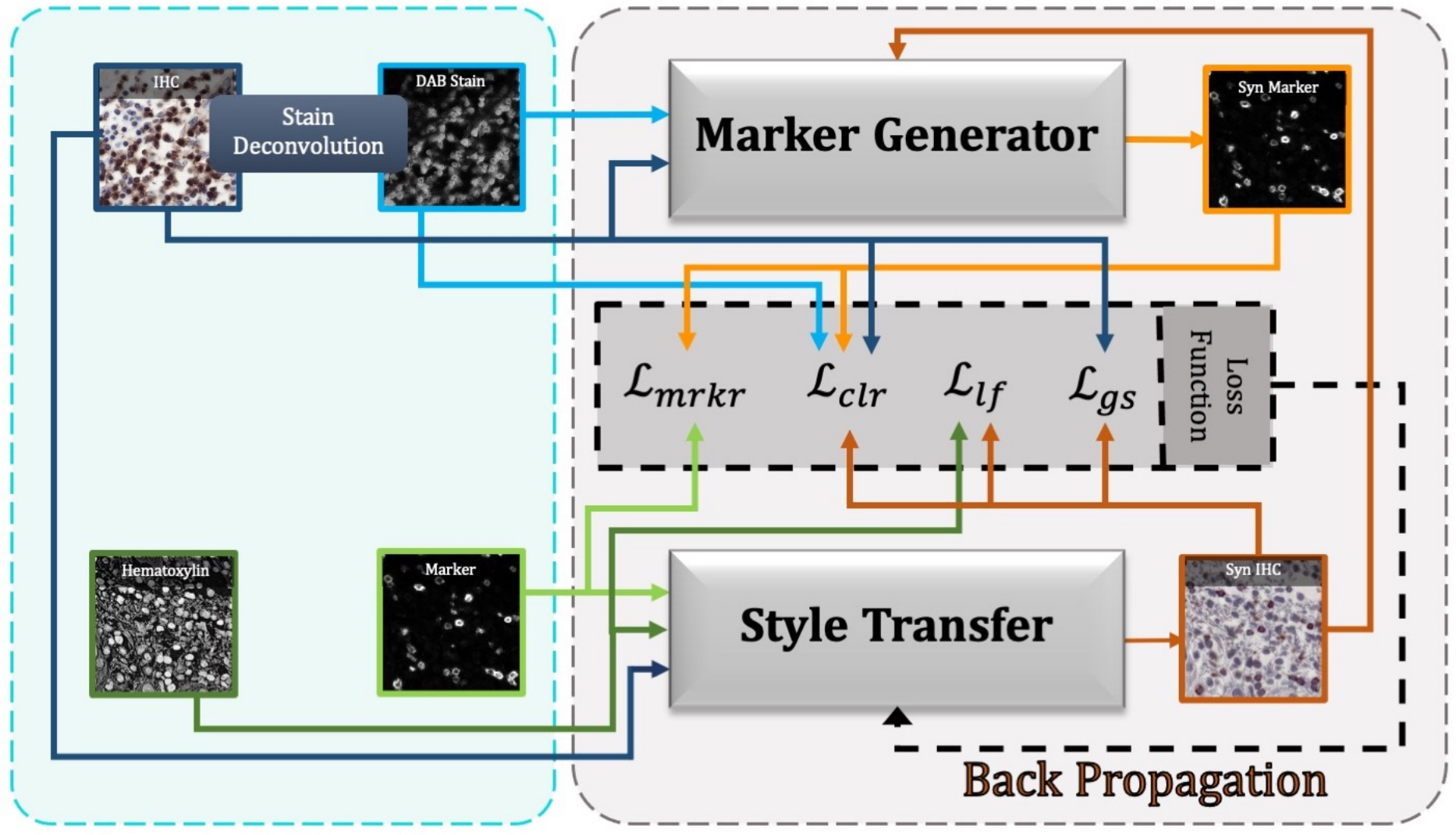}
    \caption{Overview of the lymphocytes style transfer network, where the model consists of two sub-networks. The first synthesizes the mIF marker data for the generated stylized image and is trained with the style images and their corresponding deconvolved mIF marker data. The second creates the stylized IHC image using an attention module which works on the given Hematoxylin and mIF marker images and the style and its corresponding marker images. Four loss functions including $\mathcal{L}_{lf}$ (local feature loss), $\mathcal{L}_{gs}$ (global style loss), $\mathcal{L}_{mrkr}$ (marker loss), and $\mathcal{L}_{clr}$ (color distribution loss) are calculated.}
    \label{fig:overview}
\end{figure*}

%
%
%

\begin{thebibliography}{10}
\providecommand{\url}[1]{\texttt{#1}}
\providecommand{\urlprefix}{URL }
\providecommand{\doi}[1]{https://doi.org/#1}

\bibitem{ciompi2019lymphocyte}
Ciompi, F., Jiao, Y., Laak, J.: Lymphocyte assessment hackathon {(LYSTO)}.
  \url{https://zenodo.org/record/3513571##.Yh5XJOjMJ3g} (2019)

\bibitem{ghahremani2022deep}
Ghahremani, P., Li, Y., Kaufman, A., Vanguri, R., Greenwald, N., Angelo, M.,
  Hollmann, T.J., Nadeem, S.: Deep learning-inferred multiplex
  immunofluorescence for immunohistochemical image quantification. Nature
  Machine Intelligence  \textbf{4},  401--412 (2022)

\bibitem{ghahremani2022deepliif}
Ghahremani, P., Marino, J., Dodds, R., Nadeem, S.: Deepliif: An online platform
  for quantification of clinical pathology slides. Proceedings of the IEEE/CVF
  Conference on Computer Vision and Pattern Recognition pp. 21399--21405 (2022)

\bibitem{isola2017image}
Isola, P., Zhu, J.Y., Zhou, T., Efros, A.A.: Image-to-image translation with
  conditional adversarial networks. Proceedings of the IEEE conference on
  computer vision and pattern recognition pp. 1125--1134 (2017)

\bibitem{kirillov2017unified}
Kirillov, A., He, K., Girshick, R., Doll{\'a}r, P.: A unified architecture for
  instance and semantic segmentation.
  \url{http://presentations.cocodataset.org/COCO17-Stuff-FAIR.pdf} (2017)

\bibitem{koohbanani2020nuclick}
Koohbanani, N.A., Jahanifar, M., Tajadin, N.Z., Rajpoot, N.: {NuClick:} a deep
  learning framework for interactive segmentation of microscopic images.
  Medical Image Analysis  \textbf{65},  101771 (2020)

\bibitem{liu2022bci}
Liu, S., Zhu, C., Xu, F., Jia, X., Shi, Z., Jin, M.: Bci: Breast cancer
  immunohistochemical image generation through pyramid pix2pix. arXiv preprint
  arXiv:2204.11425 (Accepted CVPR Workshop)  (2022)

\bibitem{liu2021adaattn}
Liu, S., Lin, T., He, D., Li, F., Wang, M., Li, X., Sun, Z., Li, Q., Ding, E.:
  {AdaAttN:} revisit attention mechanism in arbitrary neural style transfer.
  Proceedings of the IEEE/CVF International Conference on Computer Vision
  (ICCV) pp. 6649--6658 (2021)

\bibitem{martinez2021impartial}
Martinez, N., Sapiro, G., Tannenbaum, A., Hollmann, T.J., Nadeem, S.:
  Impartial: Partial annotations for cell instance segmentation. bioRxiv pp.
  2021--01 (2021)

\bibitem{reisenbichler2020prospective}
Reisenbichler, E.S., Han, G., Bellizzi, A., Bossuyt, V., Brock, J., Cole, K.,
  Fadare, O., Hameed, O., Hanley, K., Harrison, B.T., et~al.: Prospective
  multi-institutional evaluation of pathologist assessment of pd-l1 assays for
  patient selection in triple negative breast cancer. Modern Pathology
  \textbf{33}(9),  1746--1752 (2020)

\bibitem{ronneberger2015u}
Ronneberger, O., Fischer, P., Brox, T.: U-net: Convolutional networks for
  biomedical image segmentation. International Conference on Medical Image
  Computing and Computer-Assisted Intervention (MICCAI) pp. 234--241 (2015)

\bibitem{simonyan2014very}
Simonyan, K., Zisserman, A.: Very deep convolutional networks for large-scale
  image recognition. arXiv preprint arXiv:1409.1556  (2014)

\bibitem{stringer2021cellpose}
Stringer, C., Wang, T., Michaelos, M., Pachitariu, M.: Cellpose: a generalist
  algorithm for cellular segmentation. Nature Methods  \textbf{18}(1),
  100--106 (2021)

\bibitem{swiderska2019learning}
Swiderska-Chadaj, Z., Pinckaers, H., van Rijthoven, M., Balkenhol, M.,
  Melnikova, M., Geessink, O., Manson, Q., Sherman, M., Polonia, A., Parry, J.,
  Abubakar, M., Litjens, G., van~der Laak, J., Ciompiothers, F.: Learning to
  detect lymphocytes in immunohistochemistry with deep learning. Medical Image
  Analysis  \textbf{58},  101547 (2019)

\bibitem{zhou2018unet++}
Zhou, Z., Rahman~Siddiquee, M.M., Tajbakhsh, N., Liang, J.: Unet++: A nested
  {u-net} architecture for medical image segmentation. In: Deep Learning in
  Medical Image Analysis and Multimodal Learning for Clinical Decision Support.
  pp. 3--11. Springer (2018)

\end{thebibliography}


\end{document}